# CORNEAL THICKNESS AND ELEVATION MAPS COMPUTED FROM OPTICAL ROTARY SCANS

Sandra Franco[1], José B Almeida[1] and Manuel Parafita[2]

1. Physics Department, Universidade do Minho, Portugal.

2. Department of Surgery (Ophthalmology), Universidad de Santiago de Compostela, Spain.

Authors:

Sandra Franco, Member of Faculty

José Borges de Almeida, PhD, Member of Faculty

Manuel Parafita, PhD, Member of Faculty

Corresponding author:

Sandra Franco

Physics Department, Universidade do Minho

Campus de Gualtar

4710-057 Braga

PORTUGAL

Tel:+351253604067, Fax:+351253678981, E-mail: SFRANCO@FISICA.UMINHO.PT

# ABSTRACT

**Purpose:** Recently the authors presented a technique that allows corneal thickness measurements along any meridian from optical sections obtained using a rotary scanning system.

This paper presents three-dimensional (3-D) mapping of the corneal thickness and topography of both corneal surfaces, obtained with the rotary system.

**Method**: Corneal thickness and topography are computed from optical sections obtained by illumination with a collimated beam expanded in a fan by a small cylindrical lens. This lens is provided with motor driven rotation in order to perform automated rotary scanning of the whole cornea. Two cameras are used to capture the images of the optical sections.

**Results:** With this system it is possible to obtain measurements of the corneal thickness as well as corneal topography. Corneal thickness and elevation maps are shown.

**Conclusions:** Although still under development, this new optical system allows the measurement of the thickness of the whole cornea as well as topographical mapping of both corneal surfaces.

With the growing popularity of refractive surgery, the measurement of corneal shape, refractive power and thickness has become increasingly important. The majority of commercial videokeratographers measure the topography of the anterior corneal surface but not the posterior corneal surface and therefore provides an incomplete understanding of the total corneal power. In addition, knowledge of corneal thickness (CT) is increasingly important as it has been shown that the corneal response to different physiological and pathological conditions varies with locality.[1] Widefield or topographical determinations of CT has been used to, study pathological conditions as keratoconus,[2-7] research corneal physiology,[8, 9] and in contact lens research.[10-13] However, a number of these studies demonstrated high variability of measurements with low reproducibility in peripheral locations.[14]

One of the most common approaches for measuring corneal thickness is ultrasound pachymetry. Ultrasonic pachymeters do not require much training, are portable and produce more rapid and objective results than some other techniques.[15, 16] The technique results in contact, requiring the cornea to be anaesthetised, and measurements may be affected by hydration fluxes. Nonetheless ultrasonic pachymeters have been successfully applied to the topographical determination of CT in a variety of clinical situations.[4, 5, 17, 18] A new system for the topographical determination of CT has produced measurements with acceptable levels of reproducibility for central and peripheral measurements.[19]

Slit-lamp based optical pachymetry method, which has been for long the most commonly available optical method, allows measurements at different corneal locations on the horizontal meridian with great accuracy and reproducibility.[6, 20] New pachymetric methods based on optical technology have been recently developed and clinically applied. Confocal microscopy, videopachymetry, non-contact and contact specular microscopy and low-coherence interferometry are other techniques used in the measurement of CT. A wide review of these methods has been recently published. [21]

Nowadays, the most popular instrument for the clinical topographic determination of CT is the Orbscan Topography System (Bausch and Lomb, Rochester, NY), which uses scanning slit technology and provides topographic information of both anterior and posterior corneal surfaces as well as CT values for up to 9000 central and peripheral locations.[22-24] Several reports have pointed out that pachymetry performed by the Orbscan is reliable and reproducible. [23, 24] Other authors report errors in the calculation of corneal thickness in eyes with certain corneal opacity or those having had previous surgery [25-30]. Recently, a new system has been introduced: the Oculus Pentacam. This device is a rotating Scheimpflug camera, which scans and measures the complete cornea and anterior chamber in less than two seconds.

In this paper the authors present a new optical corneal tomographer that uses two Charge-coupled device (CCD) cameras attached to an innovative illumination system that allows widefield determination of corneal thickness as well as topographical mapping of both corneal surfaces. The optical principles and technical details of a precursor apparatus have been described. [31-33]

## MATERIAL AND METHODS

The rotary scanning system used in this study was modified from that previously described[31] with several improvements. In the present configuration it consists of an illuminator and two CCD cameras (COHU, COU 2252) provided with 55 mm telecentric lens connected to a data processing computer (Fig. 1). The illumination system comprises a quartz halogen light source, an optical fiber bundle, a collimator, a small rod shaped cylindrical lens with a diameter of 5 mm, a convex lens and an apodizing aperture slit. The cylindrical lens expands the collimated beam into a fan and is held by a mount that can automatically rotated to produce rotary scanning of the entire cornea. The fan is focused on the cornea surface by the convex lens and the light diffused from the cornea produces an optical section whose orientation follows the cylinder lens orientation. As the patient is asked to look at the light in the centre of the illumination system during the image acquisition, the illumination system is aligned with visual axis which is coincident with the rotational axis.

The diffused light is imaged by the two cameras placed at 60° with the light beam and defining with the visual axis planes perpendicular to each other. These two cameras act like a single virtual camera that can be rotated with the cylinder lens; the advantage of using two cameras is that it allows faster rotation then would be possible if a single camera had to be rotated in synchronism with the lens. The rotary scanning of the whole cornea is automatically performed simultaneously with image acquisition by the cameras.

After image acquisition, image processing reconstructs the shape and thickness of the cornea from the distorted optical sections acquired by the two cameras.

The rotary scanning of the cornea is continuous and the number of optical sections acquired is only limited by processing time. In order to get results faster and considering the processing is not yet optimized, we only acquired images from six meridians. During the image acquisition of each meridian the cylinder lens didn't rotate to ensure the position of the optical section. In this study the measurements were taken in steps of 30º but it is possible to decrease this interval at the expense of processing time. In spite of this interval, the centre of the cornea is measured very precisely because of the rotational process. The time of all image acquisition was five seconds.

After image acquisition six optical sections have been attained one from each meridian captured by the two cameras. The vertical optical section is obtained by the camera lying on the horizontal plane; the corresponding image on the vertical camera carries no information because it is reduced to a bright straight line. The situation is reversed when the optical section is horizontal, but for all other situations there is information in both images, which must be processed in order to obtain the image that would be seen by a virtual camera on a plane always normal to the optical section. The first processing step is an application of simple trigonometry to recover the virtual camera's image.

The next step is the detection of both corneal edges from the optical sections images using the method known as "adaptive thresholding" reported by Hachicha *et al*.[34] This procedure is done for each measured meridian and gives a set of points from each surface; these points are then fitted with a 4th degree polynomial and the shape of both corneal surfaces is obtained. We have also tried to use spline rather than polynomial fitting, but the processing time increased drastically.

The corneal thickness is then computed for the six meridians from the distance between the anterior and posterior edge profiles after corrections for observation angle and corneal curvature. The former of these corrections may be obtained by simple trigonometric calculus;[31] the latter is performed considering the optical magnification produced by an average curvature and could be improved by iterative processing, using pre-determined curvature at each point. The result is a thickness profile along the six meridians and it becomes possible to interpolate the thickness along all meridians and to compute a thickness map of the whole cornea. In this paper we present an elevation map for each corneal surface as well as a thickness map for one human eye measured with the rotary scanning system.

# RESULTS

Figure 2 shows an elevation map for the anterior (A) and posterior (B) surfaces. They represent the difference in height from a sphere of 7.75 mm for the anterior surface and 6.45 mm for the posterior surface.

We can observe the typical height pattern of an astigmatic surface, for both the anterior and posterior cornea, with the posterior surface demonstrating greater amplitude. Meridional deviations in elevation are more noticeable in the midperipheral region from 6 to 8 mm as represented by the colour contour map (Figure 2).

The same patterns were obtained with the Orbscan II system for the same subject also displaying greater astigmatism of the posterior surface.

Knowing the shape and position of both corneal surfaces allows the computation of the thickness of the entire cornea. Figure 3 shows a thickness map of a right eye. The values found with the rotary system are in good agreement with those found with Orbscan II for the central cornea; in the midperiphery the values found with the Orbscan II are a slightly higher than those found by us. Due to camera limitations, the images obtained have too much noise near the limbus which compromises peripheral thickness measurements.

# DISCUSSION

The authors describe a new optical system designed to provide topographical maps of both corneal surfaces and corneal pachymetry. This system allows a rotary scan of the entire cornea and the capture of optical sections from any corneal meridian. Rotary scan provides very high resolution in the central area of the cornea and is more desirable than translational scan. A cylindrical lens is used to expand a light beam in a fan and produce the corneal optical sections. The cylinder lens is provided with motor driven rotation about an axis normal to its own in order to rotate the fan of light on the cornea such that the projected line scans the whole cornea.

The preliminary findings reported in this paper are consistent with those obtained with current methods of topographic evaluation of the anterior segment. Further developments will be necessary to render the system more user friendly and to speed up the processing of different pachymetric, elevation and curvature maps similar to those currently in clinical use. The authors plan to increase the scanning speed in order to produce complete topography maps uninfluenced by eye movements; at present scanning speed is only limited by image acquisition and not by any mechanical constraints. Compared to systems currently on the market this technique allows very fast scans, 1/60 of a second is possible, avoiding problems with eye movement and equipment vibration. The present implementation does not fully exploit the technique's capabilities due to camera and software limitations.
.

Figure 1: View of the rotary scanning system.

Figure 2: Elevation maps from the anterior (A) and posterior (B) corneal surfaces.

Figure 3: Corneal thickness map.

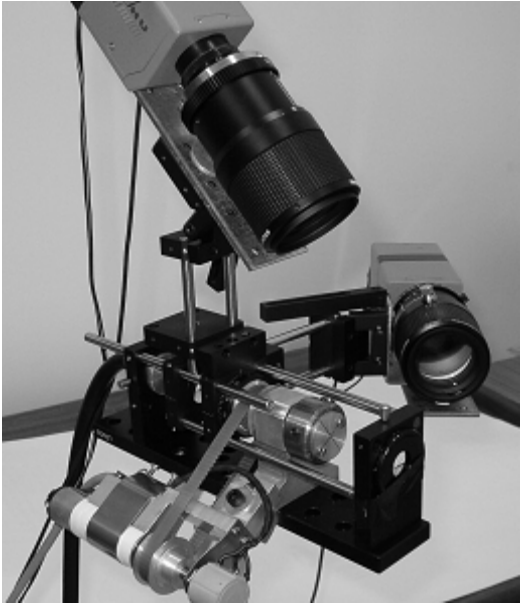

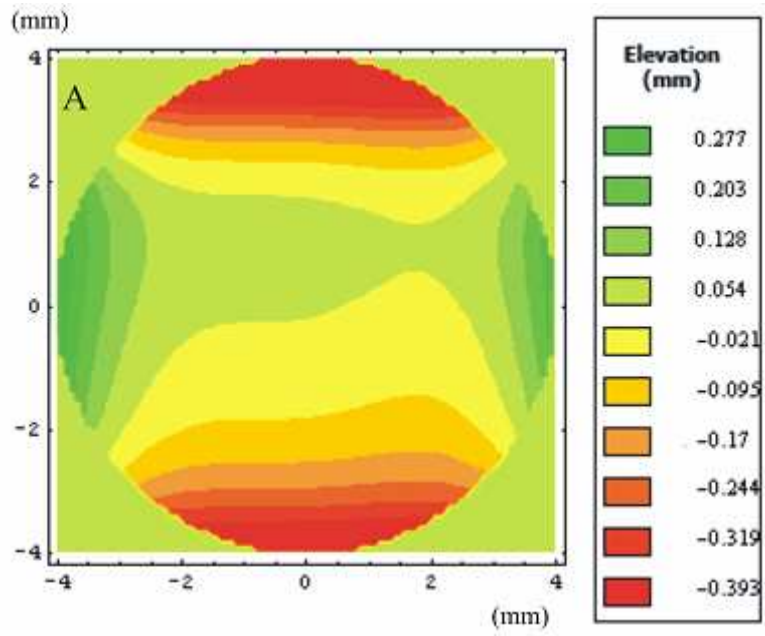
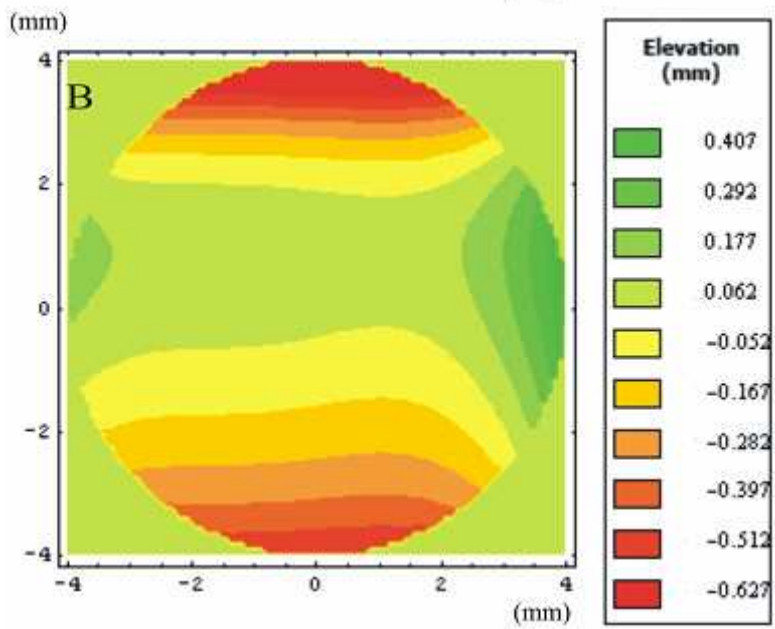

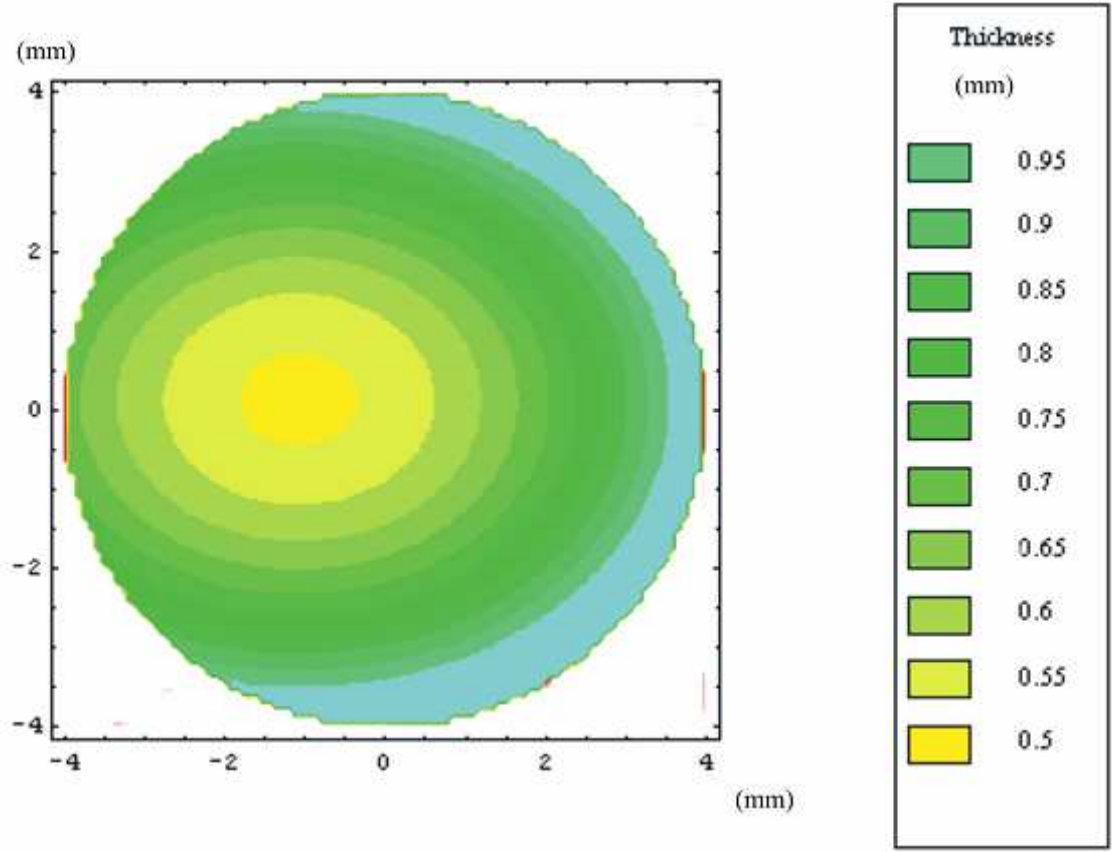